\documentclass[a4paper]{jpconf}
\usepackage{graphicx}
\begin{document}
\title{Virgo calibration and reconstruction of the gravitational wave strain during VSR1}

\author{T.~Accadia$^{12}$,
F.~Acernese$^{6ac}$,
F.~Antonucci$^{9a}$,
S.~Aoudia$^{15a}$,
K.~G.~Arun$^{11}$,
P.~Astone$^{9a}$,
G.~Ballardin$^{2}$,
F.~Barone$^{6ac}$,
M.~Barsuglia$^{1}$,
Th.~S.~Bauer$^{14a}$
% ...
}

\address{$^{1}$AstroParticule et Cosmologie (APC), CNRS: UMR7164-IN2P3-Observatoire de Paris-Universit\'e Denis Diderot-Paris 7 - CEA : DSM/IRFU}
\address{$^{2}$European Gravitational Observatory (EGO), I-56021 Cascina (Pi), Italy}
\address{$^{3}$ESPCI, CNRS,  F-75005 Paris, France}
\address{$^{4}$INFN, Sezione di Firenze, I-50019 Sesto Fiorentino$^a$; Universit\`a degli Studi di Urbino 'Carlo Bo', I-61029 Urbino$^b$, Italy}
\address{$^{5}$INFN, Sezione di Genova;  I-16146  Genova, Italy}
\address{$^{6}$INFN, sezione di Napoli $^a$; Universit\`a di Napoli 'Federico II'$^b$ Complesso Universitario di Monte S.Angelo, I-80126 Napoli; Universit\`a di Salerno, Fisciano, I-84084 Salerno$^c$, Italy}
\address{$^{7}$INFN, Sezione di Perugia$^a$; Universit\`a di Perugia$^b$, I-6123 Perugia,Italy}
\address{$^{8}$INFN, Sezione di Pisa$^a$; Universit\`a di Pisa$^b$; I-56127 Pisa; Universit\`a di Siena, I-53100 Siena$^c$, Italy}
\address{$^{9}$INFN, Sezione di Roma$^a$; Universit\`a 'La Sapienza'$^b$, I-00185  Roma, Italy}
\address{$^{10}$INFN, Sezione di Roma Tor Vergata$^a$; Universit\`a di Roma Tor Vergata$^b$; Universit\`a dell'Aquila, I-67100 L'Aquila$^c$, Italy}
\address{$^{11}$LAL, Universit\'e Paris-Sud, IN2P3/CNRS, F-91898 Orsay, France}
\address{$^{12}$Laboratoire d'Annecy-le-Vieux de Physique des Particules (LAPP),  IN2P3/CNRS, Universit\'e de Savoie, F-74941 Annecy-le-Vieux, France}
\address{$^{13}$Laboratoire des Mat\'eriaux Avanc\'es (LMA), IN2P3/CNRS, F-69622 Villeurbanne, Lyon, France}
\address{$^{14}$Nikhef, National Institute for Subatomic Physics, P.O. Box 41882, 1009 DB Amsterdam$^a$; VU University Amsterdam, De Boelelaan 1081, 1081 HV Amsterdam$^b$, The Netherlands}
\address{$^{15}$Universit\'e Nice-Sophia-Antipolis, CNRS, Observatoire de la C\^ote d'Azur, F-06304 Nice$^a$; Institut de Physique de Rennes, CNRS, Universit\'e de Rennes 1, 35042 Rennes$^b$; France}
\address{$^{16}$INFN, Gruppo Collegato di Trento$^a$ and Universit\`a di Trento$^b$,  I-38050 Povo, Trento, Italy;   INFN, Sezione di Padova$^c$ and Universit\`a di Padova$^d$, I-35131 Padova, Italy}
\address{$^{17}$IM-PAN 00-956 Warsaw$^a$; Warsaw Univ. 00-681 Warsaw$^b$; Astro. Obs. Warsaw Univ. 00-478 Warsaw$^c$; CAMK-PAN 00-716 Warsaw$^d$; Bia\l ystok Univ. 15-424 Bial\ ystok$^e$; IPJ 05-400 \'Swierk-Otwock$^f$; Inst. of Astronomy 65-265 Zielona G\'ora$^g$,  Poland}
\address{$^{*}$University of Birmingham, Birmingham, B15 2TT, United Kingdom }
\address{$^{**}$University of Glasgow, Glasgow, G12 8QQ, United Kingdom }


\ead{rollandl@in2p3.fr}

\begin{abstract}
Virgo is a kilometer-length interferometer for gravitational waves detection located near Pisa.
Its first science run, VSR1, occured from May to October 2007.
The aims of the calibration are to measure the detector sensitivity
and to reconstruct the time series of the gravitational wave strain $h(t)$. 

The absolute length calibration is based on an original non-linear reconstruction of the differential arm
length variations in free swinging Michelson configurations. It uses the laser wavelength as length standard.
This method is used to calibrate the frequency dependent response of the Virgo mirror actuators
and derive the detector in-loop response and sensitivity within $\sim5\%$. 

The principle of the strain reconstruction is highlighted and the $h(t)$ systematic errors are estimated.
A photon calibrator is used to check the sign of $h(t)$.
The reconstructed $h(t)$ during VSR1 is valid from 10~Hz up to 10~kHz 
with systematic errors estimated to 6\% in amplitude.
The phase error is estimated to be 70~mrad below 1.9~kHz and 6\,$\mathrm{\mu s}$ above.
\end{abstract}

\section{Introduction}
The Virgo detector~\cite{bib:VirgoDetector} is a kilometer-length interferometer (ITF) 
located near Pisa (Italy). It is designed to search for gravitational waves (GW) 
in the frequency range from 10~Hz to a few~kHz.
Expected astrophysical sources of detectable GWs are compact objects such as neutrons stars or black holes.

The Virgo first science run (VSR1) was performed from May 18th to October 1st 2007
with a sensitivity close to its nominal one, and in coincidence to the end of the
fifth science run (S5) of the three LIGO detectors~\cite{bib:LigoDetectors}.
In order to reduce the false detection rate and improve the estimation of the parameters of potential sources,
the data of all the detectors are used together to search for a GW signal.

\begin{figure}[ht]
\includegraphics[width=34pc]{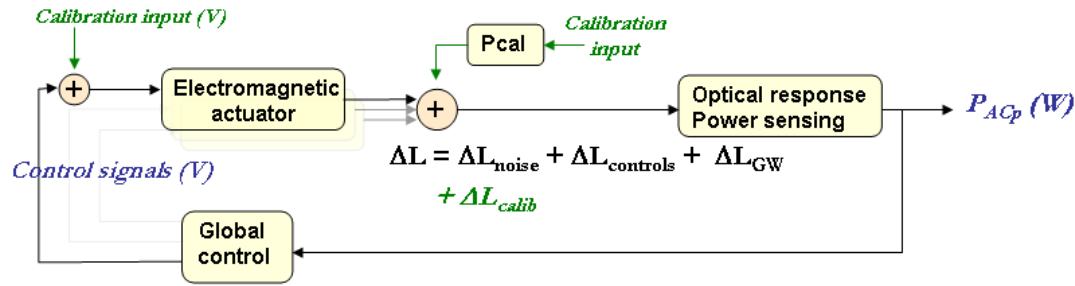}
\caption{
\label{fig:LongitudinalLoop}
Synopsis of the Virgo longitudinal control loop.
A variation of the differential length of the arms of the ITF, $\Delta L$, is sensed through the
optical response which converts the length variation into laser power variation
at the ITF output port. The power impinging on photodiodes is read and sampled at 20~kHz
in the channel $P_{ACp}$ (W).
This channel is used as error signal to compute the control signals (V) sent to the different mirror actuators.\\
Calibration signals can be injected into the loop to force differential length variations $\Delta L_{calib}$. 
The standard path is through the electromagnetic actuators.
An auxiliary setup, the photon calibrator (pcal), can also be used (see section~\ref{lab:Pcal}).
}
\end{figure}

The purpose of the Virgo calibration is to measure the parameters needed
(i) to estimate the ITF sensitivity to GW strain as a function of frequency, $\tilde{h}(f)$, and
(ii) to reconstruct from the ITF data the amplitude $h(t)$ of the GW strain signal.
To achieve optimum sensitivity, the positions of the different mirrors are controlled~\cite{bib:LockAcquisition} 
to have, in particular, beam resonance in the cavities, destructive interference at the ITF output port
and to compensate for environmental noise.
The controls would also attenuate the effect of a passing GW below a few hundreds hertz. 
A synopsis of the longitudinal control loop
and its components is given in figure~\ref{fig:LongitudinalLoop}.
The effects of the longitudinal controls have thus to be precisely calibrated in the frequency domain.
Above a few hundreds hertz, the mirrors behave as free falling masses in the longitudinal direction.
The main effect of a passing GW would then be a frequency-dependent variation of the power at the ITF output.
characterized by the ITF optical response.
Finally, the readout electronics of the output power and its timing precision have also to be calibrated.

In this paper, the main method used to get an absolute length calibration is first described 
and the mirror actuator calibration results are highlighted. 
The way the Virgo sensitivity is measured is then shown.
In the last section, the principle of the h(t) reconstruction is described as well as some
measurements performed to estimate the errors on the h(t) signal.

\section{Absolute length measurement for mirror actuation calibration}
\label{lab:FreeMich}
The ITF calibration is based on absolute length measurements. 
The displacement induced by the mirror actuators is reconstructed using the ITF as
a simple Michelson. The non-linear method used to reconstruct the arm differential motion
in the freely swinging Michelson with passing fringes is described. 
The laser wavelength is used as length standard.

Mirrors of the Virgo ITF are mis-aligned to get a simple Michelson configuration.
Different configurations are used to calibrate the beam-splitter and arm mirrors actuations
(comprising the beam-splitter, one end mirror and the input mirror in the other arm).
The typical results of the mirror actuation calibration are given.

\subsection{Absolute length measurement}
\label{lab:FreeMichRec}

\begin{figure}[h]
\begin{minipage}{17pc}
\includegraphics[width=17pc]{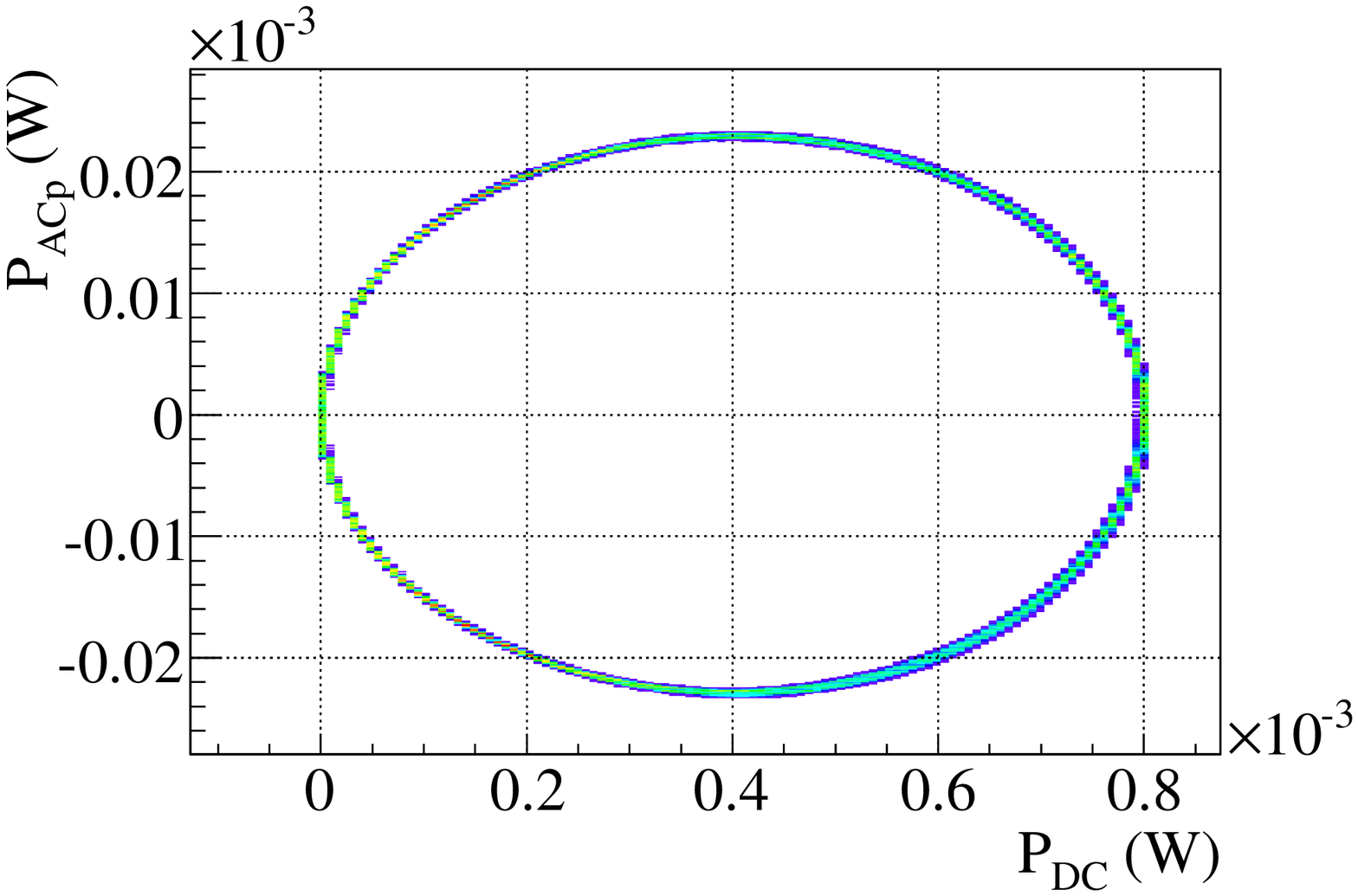}
\caption{\label{fig:FreeMichEllipse}
         $\mathcal{P}_{AC}$ vs $\mathcal{P}_{DC}$ measured in free swinging Michelson data
	(configuration comprising the beam-splitter and the two arm input mirrors).
}
\end{minipage}\hspace{2pc}%
\begin{minipage}{19pc}
\includegraphics[width=16pc]{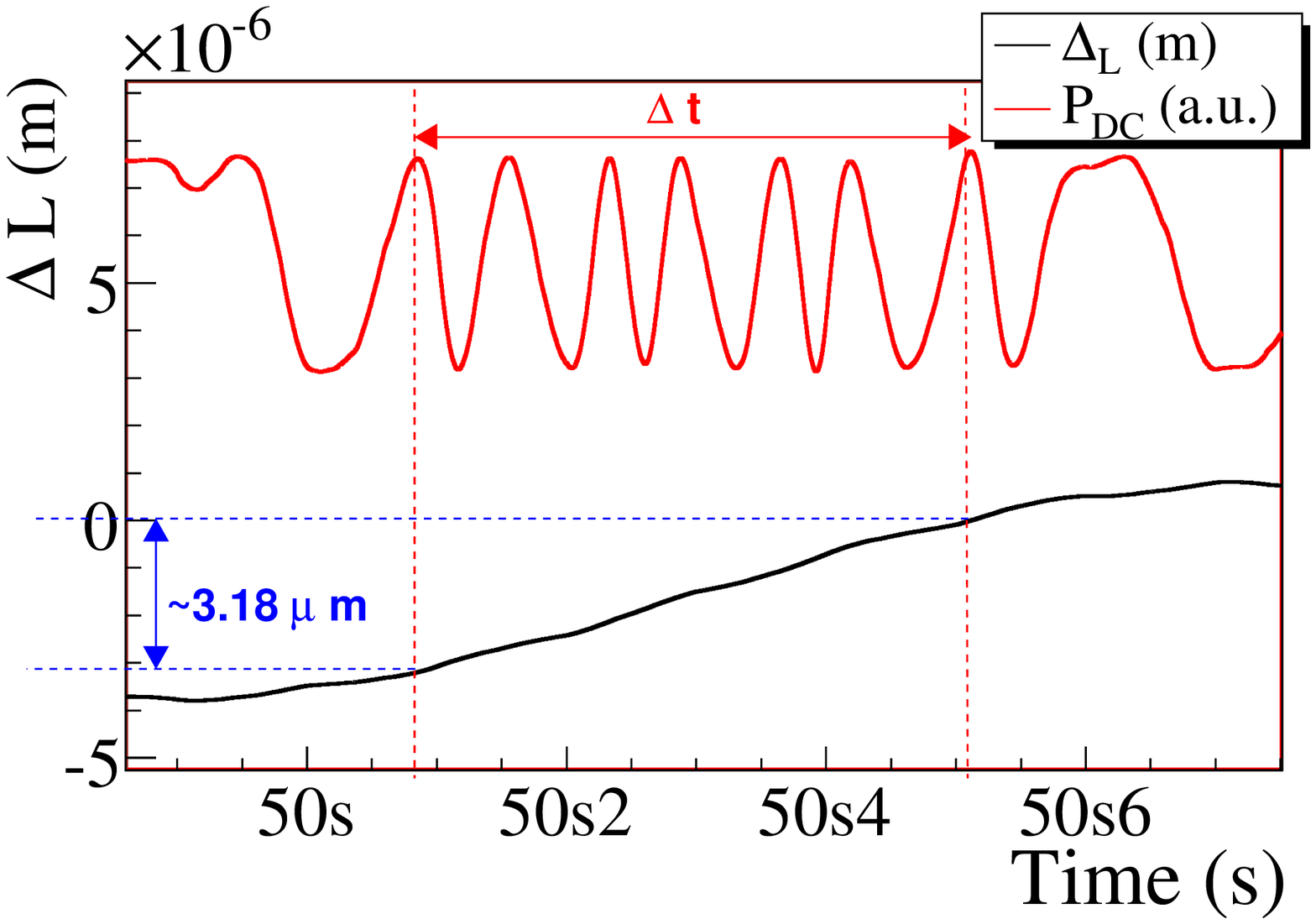}
\caption{\label{fig:FringeCounting}
	Red curve (top): measured time series of $\mathcal{P}_{DC}$ in free swinging Michelson data, in arbitrary units.
	Black curve (bottom): reconstructed time series of $\Delta L$, in meters.
}
\end{minipage} 
\end{figure}

In a simple Michelson ITF, the phase difference $\Delta \Phi$ between  the two interfering beams
is function of the differential arm length $\Delta L$, using the laser wavelength $\lambda=1064\,\mathrm{nm}$ as standard:
$\Delta \Phi(t) \,= \, \frac{4\pi}{\lambda} \Delta L(t)$\label{eqn:FreeMichPhase}. \\
In a simple Michelson ITF with frontal phase-modulation, 
the power of the output beam ($\mathcal{P}_{DC}$) as well as the demodulated power ($\mathcal{P}_{AC}$) signals
are functions of the phase difference $\Delta\Phi$ between the two interfering beams:
$\mathcal{P}_{DC} \,=\,\beta(1-\gamma\cos(\Delta\Phi))$ and
$\mathcal{P}_{AC} \,=\, \alpha \sin(\Delta\Phi)$, 
where $\alpha$ and $\beta$ are proportional to the laser power 
and $\gamma$ is proportional to the ITF contrast.
Therefore, in the ($\mathcal{P}_{DC},\mathcal{P}_{AC}$) plane, 
the signal follows an ellipse as shown in figure~\ref{fig:FreeMichEllipse}.

The measurement of the differential arm length $\Delta L$ requires a non-linear reconstruction.
The ellipse is fitted using the method described in~\cite{bib:HalirFlusser}. 
The fit gives the ellipse center position (theoretically ($\beta$,0)) and the axis widths 
(theoretically ($\beta\gamma,\alpha$)). 
The variations of the parameters $\alpha$, $\beta$ and $\gamma$
are monitored and found to be of the order of 0.5\% during a few-minute dataset.
$\Delta\Phi'$ can then be estimated directly from the ITF signals 
as the angle between the ellipse axis along DC and the line from the ellipse center 
to the current point position ($\mathcal{P}_{DC},\mathcal{P}_{AC}$).
Using a suitable ellipse tour counting, the right number of $2\pi$ is added to $\Delta\Phi'$
to recover completely the angle $\Delta\Phi$.
The differential arm length $\Delta L_{rec}$ is then computed from equation~\ref{eqn:FreeMichPhase}.\\
Figure~\ref{fig:FringeCounting} is an illustration of the method.
In the window $\Delta t$, 6~interfringes passed on the DC signal:
It indicates a differential arm elongation of $6\times \frac{\lambda}{2} = 3.19\,\mathrm{\mu m}$.
In the same window, the reconstructed $\Delta L$ varies by $\sim3.18\,\mathrm{\mu m}$.

\subsection{Mirror actuator calibration}
\begin{figure}[ht]
\includegraphics[width=32pc]{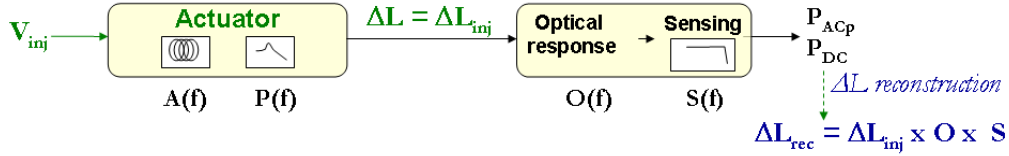}
\caption{Signal processing in the free swinging Michelson data used for the mirror actuation calibration.
	See text for details.
\label{fig:FreeMichelsonChain}
}
\end{figure}

The Virgo electromagnetic actuators are used to induce a longitudinal motion to the suspended mirrors
from a voltage signal. As shown in figure~\ref{fig:FreeMichelsonChain}, the signals pass through 
some electronics and  pairs of coil-magnet ($A$) and the pendulum mechanical response ($P$). 
Their calibration consists in measuring the transfer function between the input voltage $V_{inj}$ 
and the induced motion $\Delta L_{inj}$. 
This is done setting the ITF in free swinging Michelson configurations: 
sine signals $V_{inj}$ are injected to the actuators and the motion $\Delta L_{rec}$ is resconstructed using the method described in section~\ref{lab:FreeMichRec}. 
The complete response of the ITF is shown in figure~\ref{fig:FreeMichelsonChain}: the reconstructed $\Delta L_{rec}$ 
has to be corrected for some effects to get the true induced mirror motion $\Delta L_{inj}$, 
mainly some delays due to the light propagation time in the arms ($O$) and due to the output power readout electronics ($S$). 

Thirteen sine signals have been injected from 5~Hz to 1.~kHz. Such injections were done at different amplitudes
in order to check the linearity of the measurements. They were also performed every two weeks during VSR1 in order
to monitor their time stability. 
The mirror actuation responses $\Delta L_{inj}/V_{inj}$ are measured with a statistical error better than 1\%.
A regular monitoring of the actuation has shown $\sim2\%$ variations and different measurements 
had up to $\sim2\%$ offsets from the standard measurements.
Systematic errors of the order of 4\% on the modulus and 50~mrad on the phase have been derived.

\section{Virgo ITF response and sensitivity measurements}
The first goal of the calibration is to measure the Virgo sensitivity.
It is based on the knowledge of a mirror actuator response calibrated as shown in section~\ref{lab:FreeMich}.
The ITF response in closed-loop is first measured and then used to estimated the ITF sensitivity.

\subsection{Virgo ITF response}
The ITF response, in W/m,  is the the transfer function from the differential arm length variation $\Delta L$
to the ITF output power $\mathcal{P}_{ACp}$. 
When the Virgo ITF is in closed-loop (see figure~\ref{fig:LongitudinalLoop}), 
some white noise $V_{inj}$ is injected up to 1.5~kHz through 
an end mirror actuator. Using the actuator calibration, the induced $\Delta L_{inj}$ is derived.
The transfer function $\Delta L_{inj} / \mathcal{P}_{ACp}$ (inverse of the ITF response) 
is then measured up to $\sim 1.5$~kHz.

The longitudinal controls do not act above a few hundred Hz: the ITF response is only the
optical response and the output power sensing. A simple model is thus fit to the measured transfer function
from 800~Hz to 1~kHz.
The optical response is modeled by a simple pole at 500~Hz
due to the arm cavities with finesse 50. 
The finesse of the arm cavities changes over time by as much as $3\%$.
These variations are ignored at this stage.
The gain of the optical response is let free in the fit.
The measurements of the output power sensing are not described in this paper: 
the fitted model of the sensing lies within 2\% in modulus and 20~mrad in phase from the measured response up to 6~kHz.

An example of a measured inverse ITF response during VSR1 is given in figure~\ref{fig:ITFresponse}.

\subsection{Virgo sensitivity}
\begin{figure}[ht]
\begin{minipage}{18pc}
\includegraphics[width=17pc]{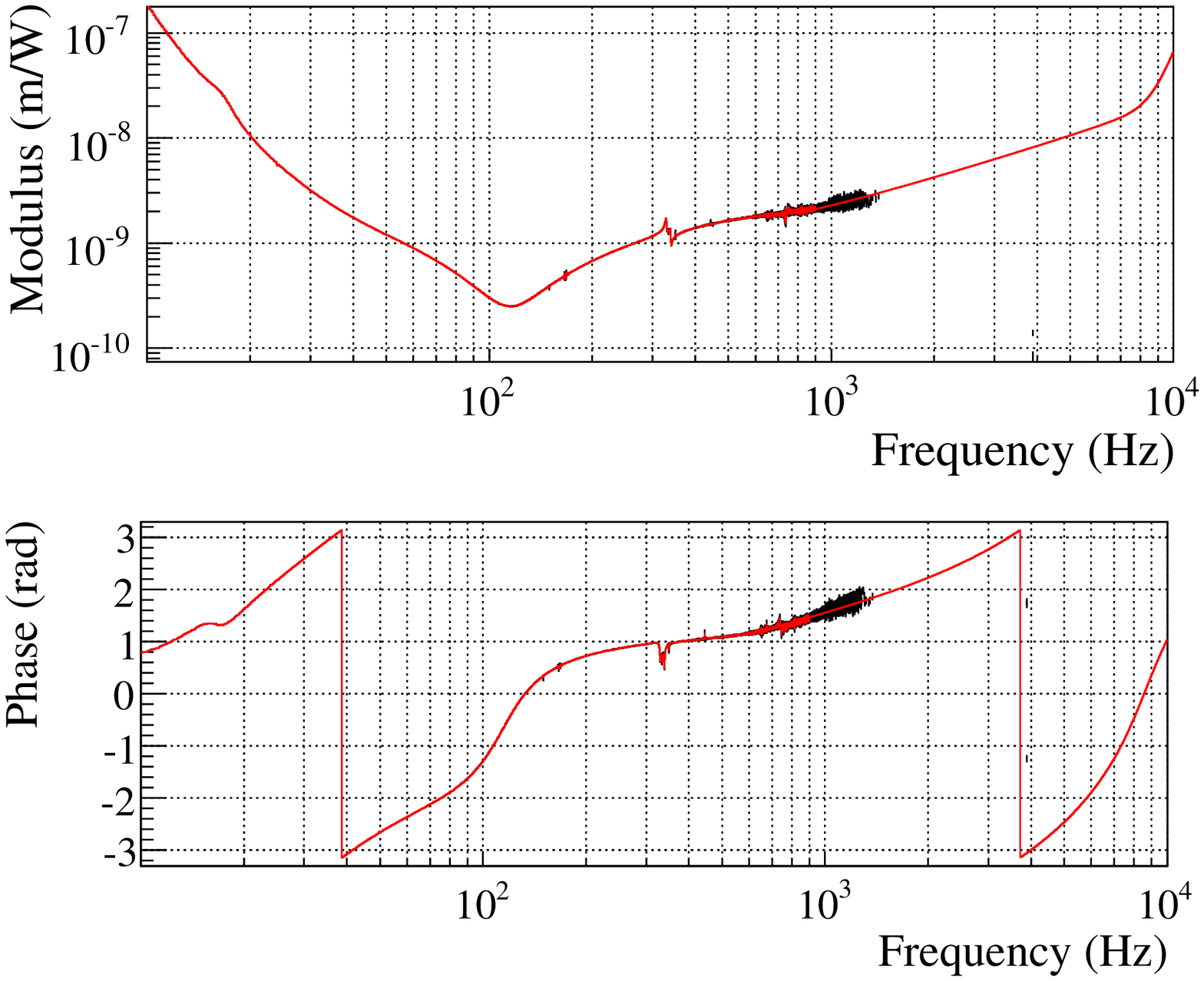}
\caption{\label{fig:ITFresponse}
	Virgo ITF response measured in September 2007.
	Black: measured transfer function up to 1.5~kHz (with increasing statistical uncertainties above a few 100's Hz).
	Red: extrapolated transfer function.
}
\end{minipage}\hspace{2.0pc}%
\begin{minipage}{18pc}
\includegraphics[width=18pc]{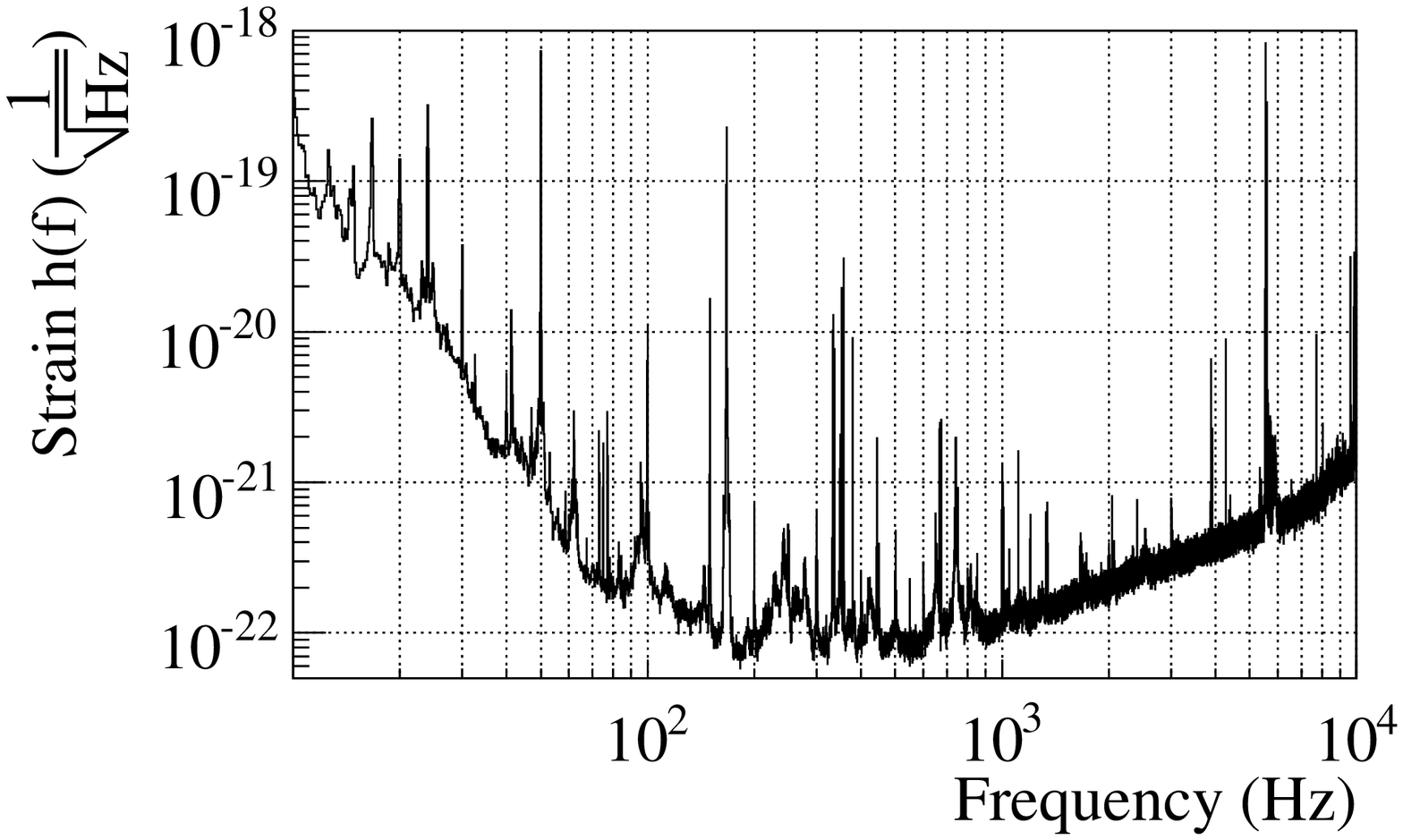}
\caption{\label{fig:ITFsensitivity}
	Virgo sensitivity measured in September 2007.
}
\end{minipage}
\end{figure}

The Virgo sensitivity is measured from the ITF response in closed-loop, in m/W, and from
the spectrum of the output signal $\mathcal{P}_{ACp}$, in W: they are multiplied
to get the sensitivity in meters. It is then divided by the Virgo arm length, 3~km,
to get the sensitivity in strain as shown in figure~\ref{fig:ITFsensitivity}.
The sensitivity is measured a few minutes after the ITF response in order to avoid
possible variations of the optical gain.

Note that the sensitivity is measured with systematic errors of the order of 4\% coming
from the mirror actuator calibration below 1~kHz. The errors above 1~kHz are higher (5--10\%) due to 
(i) the use of a fixed cavity finesse, 
(ii) the power sensing model errors and 
(iii) the fact that the controls were not totally negligible in the fitted frequency range during VSR1.

\section{h(t) strain reconstruction}
The second aim of the calibration process is the reconstruction of the GW strain time series $h(t)$.
The principle of the reconstruction used for the VSR1 data is described.
Some estimations of the errors and the sign of $h(t)$ are then given.

\subsection{Reconstruction principle}
\begin{figure}[ht]
\begin{minipage}{16pc}
\includegraphics[width=16pc]{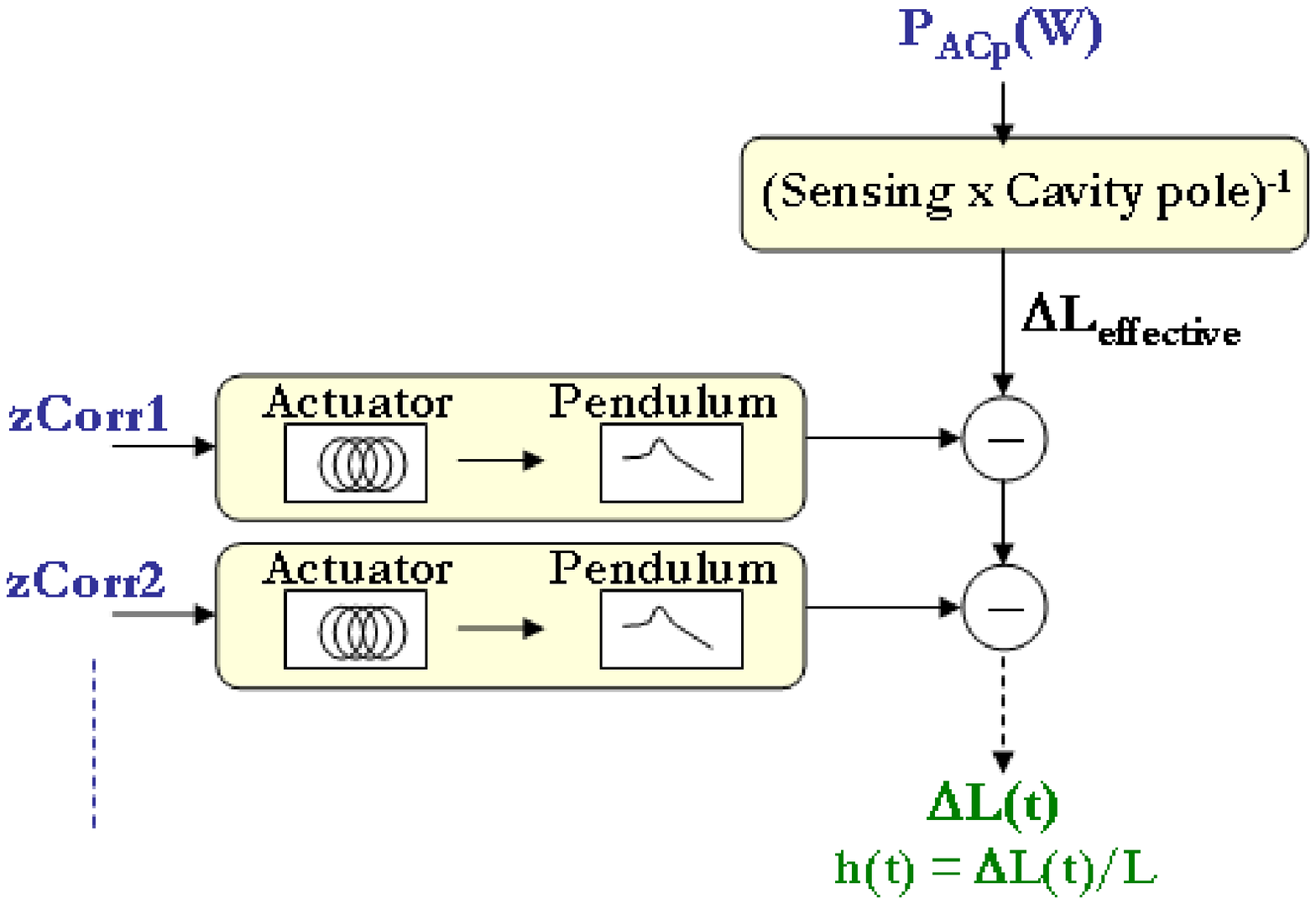}
\caption{\label{fig:HrecPrinciple}
	Synopsis of the reconstruction of $h(t)$ from the ITF output power $\mathcal{P}_{ACp}$
	and the longitudinal correction signals sent to the mirror actuators ($zCorr$ channels).
	See text for details.
}
\end{minipage}\hspace{2.0pc}%
\begin{minipage}{20pc}
\includegraphics[width=20pc]{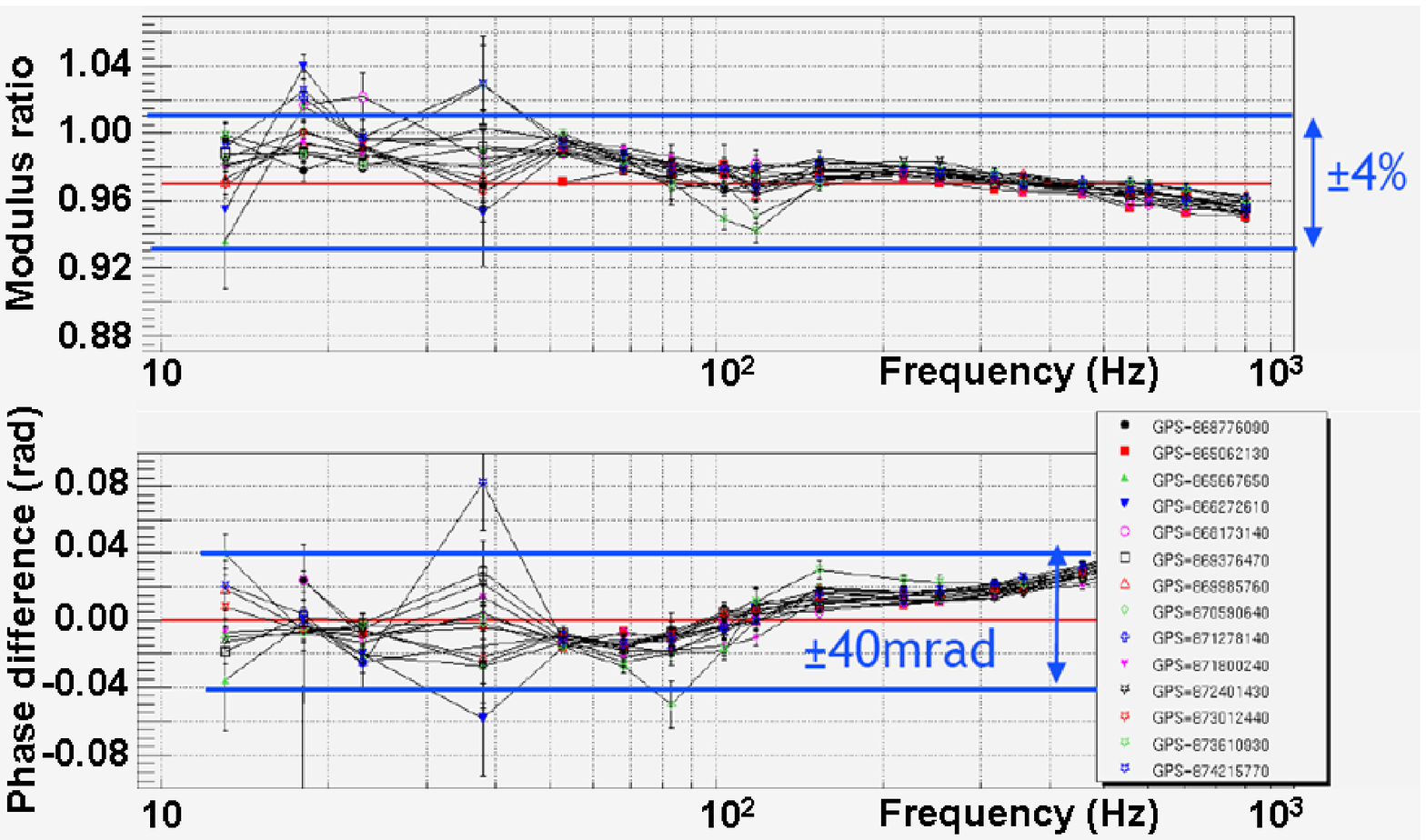}
\caption{\label{fig:HrecCheck}
	Top curve: ratio between the reconstructed amplitude of the lines in $h(t)$
	and the expected injected amplitude.
	Bottom curve: phase difference for the two signals.
	The difference symbols correspond to different measurements performed once per week during VSR1.
}
\end{minipage}
\end{figure}

The principle of the reconstruction of $h(t)$ from the ITF output power and the
mirror longitudinal controls is given in figure~\ref{fig:HrecPrinciple}.
The effective arm differential length variation $\Delta L_{effective}$ is reconstructed
in the frequency domain from the output power $\mathcal{P}_{ACp}$ 
and the calibrated optical response and power sensing. 
It contains the contribution from the noise, from the controls and possibly from the GWs. 
The contributions from the controls are reconstructed in the frequency
domain from the correction signals sent to the mirror actuators knowing the actuator responses.
They are then subtracted from $\Delta L_{effective}$ to get the
final $\Delta L$ with the noise and GW contributions only. The result is then divided
by 3~km and put back in the time domain to get the strain $h(t)$. 
The contribution from the power line (50~Hz and harmonics) is finally subtracted.

\subsection{Error estimation using mirror actuators}
In order to estimate the errors on the reconstructed strain $h(t)$,
some datasets with sine signals $V_{inj}$ injected to an out-of-loop mirror (an input mirror)
were taken every week during VSR1. 
It corresponds to an injected strain $h_{inj} = \Delta L_{inj}/(3000\,\mathrm{m})$.
The transfer function of the reconstructed over injected strain $h/h_{inj}$ is then measured.
This transfer function is expected to be flat: at 0 in phase and at 0.97 in modulus
(since the motion of the input mirror changes both the arm length and the Michelson length, 
the optical response of Virgo is 3\% lower for an input mirror than for an end mirror).

The differential arm length variations $\Delta L_{inj}$ induced by the injected signal
is derived from the mirror actuation calibration, with errors of the order
of 4\% in modulus and 50~mrad in phase. 

The transfer function measured every week during VSR1 is shown in figure~\ref{fig:HrecCheck}.
The observed time variations at a given frequency are compatible with statistical errors.
The difference between the measurements and the expected values gives an estimation
of the systematic errors on the reconstructed strain $h(t)$. 
They are of the order of 4\% in modulus and 50~mrad in phase below 1~kHz. 
This is compatible with the estimated systematic errors of the mirror actuation calibration
used in the $h(t)$ reconstruction.

\subsection{''Photon calibrator'' and sign of h(t)}
\label{lab:Pcal}
As in GEO~\cite{bib:PCalGEO} and LIGO~\cite{bib:PCalLIGO,bib:ProceedingRick}, a photon calibrator (pcal) was setup in Virgo during VSR1. 
It is used as an independent mirror actuator: the radiation pressure of an auxiliary laser is used to
displace one input mirror. Its main use during VSR1 has been to check the sign of $h(t)$,
defined in common with LIGO as $h = \frac{L_x - L_y}{L}$ where, in Virgo, $x$ and $y$ are the north and west arms respectively.

Sine signals injections were performed with the Virgo pcal, corresponding to an expected signal $h_{inj}$.
The phase between the reconstructed $h$ and the expected one $h_{inj}$ has been measured below 1~kHz.
It is expected to be~0. The phase at low frequency converges to 0 within 15~mrad: this proves that the sign of
the reconstructed $h(t)$ is correct.

Note that the pcal has been used also to cross-check the mirror actuation standard calibration.
It agrees within the large pcal systematic errors of 20\%.

\section{Conclusion}
The Virgo standard calibration is based on a non-linear reconstruction of 
the error signal in free swinging Michelson configurations.
It relies on the laser wavelength as length standard. 
It has been used to calibrate the mirror electromagnetic actuators
within 4\% in modulus and 50~mrad in phase.

The Virgo photon calibrator has been mainly used to check the sign of $h(t)$. 

The principle of the reconstruction of the strain $h(t)$ applied on the VSR1 data has been given
along with some estimations of the errors. Some other sources of errors must be taken into account, 
especially from the ITF output power sensing and from some timing issues.
Taking them into account, the reconstructed $h(t)$ during VSR1 is valid from 10~Hz up to the Nyquist frequency of the
channel (2048~Hz, 8192~Hz or 10000~Hz) with systematic errors estimated to 6\% in amplitude.
The phase error is estimated to be 70~mrad below 1.9~kHz and 6\,$\mathrm{\mu s}$ above.


\end{document}